# Trapped charge driven degradation of perovskite solar cells

Namyoung Ahn[1,2][†], Kwisung Kwak[1,2][†], Min Seok Jang[1], Heetae Yoon[1,2], Byung Yang Lee[3], Jong-Kwon Lee[1], Peter V. Pikhitsa[1], Junseop Byun[1,2], and Mansoo Choi[1,2]*

[1]*Global Frontier Center for Multiscale Energy Systems, Seoul National University, Seoul 08826, Korea*
[2]*Department of Mechanical and Aerospace Engineering, Seoul National University, Seoul 08826, Korea*
[3]*Department of Mechanical Engineering, Korea University, Seoul 02841, Korea*

[†]*These authors contributed equally to this work.*
*To whom correspondence should be addressed. E-mail: mchoi@snu.ac.kr*

**Metal halide perovskite solar cells**[1,2,3] **have shown unprecedent performance increase up to 22% efficiency**[4] **and are now considered not only as a low-cost alternative to commercialized solar cells**[5,6]**, but also as a new functional cell with flexible applications**[7,8]**. However, stability of the perovskite materials under ambient condition has not been solved yet and is the most important obstacle before commercialization and its mechanism has been elusive. Here, we reveal the fundamental mechanism for irreversible degradation of perovskite materials in which trapped charges regardless of polarity play a decisive role. A novel experimental set-up utilizing the deposition of ions with different polarity via a corona discharger reveals that perovskite materials degrade irreversibly through grain boundaries only when moisture is introduced on charge trapped perovskite surface, which indicates that the moisture induced irreversible dissociation of perovskite materials is triggered by trapped charges. This not only explains why the degradation of perovskite solar cells begins from different side of interface contacting hole or electron extracting layer depending on the use of different charge extraction layer, but also why the light soaking on the perovskite materials consistently induces irreversible degradation in the presence of moisture. The Kelvin Probe Force Microscopy measurements confirm that charges are trapped preferentially on the grain boundaries even under uniform deposition of ions or uniform illumination of light supporting the role of trapped charges in the degradation through grain boundaries. We also identified the synergetic effect of oxygen on the process of moisture induced degradation, which is directly related to the revealed mechanism of trapped charge driven degradation. The deprotonation** **of organic cations by trapped charge induced local electric field is attributed to the initiation of irreversible decomposition.**

Metal halide perovskites have been attracting worldwide interest and the power conversion efficiency (PCE) of perovskite based solar cells has already exceeded 22%. Their long-term stability issue is the most pressing problem for commercialization[9]. It is well known that perovskite materials are vulnerable to the exposure of humidity and light[10,11,12]. Though many efforts to encapsulate the devices for preventing direct contact to humidity have been attempted, it was not successful to obtain long-term stability comparable to commercial photovoltaic devices[13]. Various factors that can affect the stability have been investigated from the viewpoints of chemical structure[14], electrical stress[15], hydrated states[10,11,16] and heat[17]. However, the degradation mechanism is still unclear how a fully fabricated device deteriorates rapidly even though its perovskite layer is tightly-covered by the hole transport material(HTM) and the back electrode. It is also elusive why light soaking causes irreversible degradation of perovskite materials in the presence of moisture while the moisture in the dark condition only induces reversible hydration of perovskite materials[11,15].

The device structure and the choice of charge extraction materials also influence the stability. The use of inorganic charge extraction layers was reported to enhance stability[13,18]. Other studies have focused on eliminating the pathway of water vapor infiltration into the perovskite film by coating carbon-based materials, polymers, and hydrophobic materials on the top surface of the perovskite film[12,19]. These approaches provide the device lifespan longer than the case without these materials, but not long enough to ensure long-term stability. Such approaches also occasionally sacrifice the photovoltaic performance. Especially, the devices employing titanium dioxide ($TiO_2$) as electron transport layer(ETL) are rapidly degraded under light soaking[20] even though they are highly efficient in energy conversion and exhibit the world's best efficiency.

In the present study, we demonstrate that the charges trapped at the interface between perovskite and charge extraction materials are responsible for the irreversible degradation due to moisture. The elusive experimental puzzles both on the occurrence of degradation beginning from different side depending on different charge extraction material and the role of light soaking for irreversible degradation will be clearly explained from the present concept of the trapped charges later. To dig into this charge driven degradation mechanism, we investigated controlled stability experiments both for commonly used $CH_3NH_3PbI_3$ ($MAPbI_3$) that is known to form structurally distorted tetragonal crystals and a new mixed perovskite material having more enhanced structural stability. The crystal structure of perovskite can become more stable by increasing the tolerance factor close to unity by incorporating other organic cation and halide anion with different ion sizes[21] (See Supplementary Fig. 1). Addition of formamidinuim (FA) cation and bromide (Br) anion could not only structurally stabilize the perovskite materials by increasing tolerance factor towards one for inducing cubic crystals, but also enhance their photovoltaic performance by broadening absorption spectrum[5,6]. Although several studies on mixed cation and/or halide anion system of $MA_xFA_{1-x}PbI_yBr_{3-y}$ have been suggested[5,6,22], they focused on the performance aspect, not the stability aspect. Therefore, a new effort is required to develop new composition perovskite ensuring both stability and high performance. In this work, we developed a new mixed $MA_{0.6}FA_{0.4}PbI_{2.9}Br_{0.1}$ perovskite ensuring both high performance and stability via lewis-base adduct method[23] (see the detailed methods and characterization in Supplementary Information: Optimization of mixed $MA_xFA_{1-x}PbI_yBr_{3-y}$). Later, it will be shown that this new mixed perovskite still degrades although its degradation speed is slower than the case of conventional MAPbI$_3$ and the irreversible degradation of both perovskites is triggered by trapped charges.

First, we examined how solar cell degradation behavior becomes different depending on different charge extraction layers, for example, C$_{60}$ and compact TiO$_2$ (see the detailed process for solar cell fabrication in Supplementary Information : Solar cell fabrication ). Figure 1a and b show J-V curves for MA$_{0.6}$FA$_{0.4}$PbI$_{2.9}$Br$_{0.1}$ perovskite on the C$_{60}$ ETL (Fig.1a) and compact TiO$_2$ layer (Fig.1b), respectively. Fig.1a shows J-V curve for our mixed perovskite on 35 nm thick dense C$_{60}$ layer deposited by thermal evaporation, which demonstrates hysteresis-less performance of the best PCE of 20.2%. The best PCE value was averaged from the J-V curves of forward and reverse scan, which is in agreement with 20.2% of steady-state efficiency shown in Supplementary Fig 4b. The integrated Jsc estimated from external quantum efficiency (EQE) was also well-matched with the measured Jsc as shown in Supplementary Fig 4c. Histograms of the short-circuit current (Jsc), the open-circuit voltage (Voc), the fill factor (FF) and the efficiency of 47 cells are shown in Supplementary Fig 4d, e, f, g. The photovoltaic characteristics of these cells were highly reproducible with a small standard deviation, and the average values are Jsc = 24.34 mA/cm2, Voc = 1.058V, FF = 0.743, and PCE = 19.12%, respectively. This would be the best performance of low-temperature processed perovskite solar cells without using mesoporous TiO$_2$. On the other hand, the case coated on 40 nm thick compact TiO$_2$ showed a large hysteresis with PCE of 16.9% on the reverse scan and 9.0% on the forward scan, which is consistent with previous studies that also showed a large hysteresis for compact TiO$_2$ based devices[24].

As shown in Fig. 1c, d, non-encapsulated $C_{60}$ based cell shows much more stable performance under one sun illumination but still degrades while compact $TiO_2$ based non-encapsulated cell completely died only after 6 hours. To examine the detailed evolution of degradation, we investigated how the cross sectional morphology of the $C_{60}$ and $TiO_2$ based devices would evolve under illumination via the focused ion beam (FIB) assisted scanning electron microscope (SEM) images shown in Fig 1e, f. Consistent with the PCE measurement results, the SEM images clearly confirm that the $C_{60}$-based devices are much slowly degraded compared to the $TiO_2$-based cells. Strikingly, they showed different degradation patterns, namely, different side of degradation beginning where the degradation is initiated depending on different ETLs. Such degradation pattern is the same for conventional MAPbI3 perovskite (see Supplementary Fig. 5). Since the reactants that can decompose perovskite materials could infiltrate from the thin metal electrode rather than from the thick ITO glass, it would be expected that the degradation should be initiated at the interface closer to the thin Au metal electrode. However, the perovskite films of $TiO_2$-based devices began to be decomposed at the interface adjacent to the compact $TiO_2$ layer near FTO glass (as shown in Fig. 1f). For those of $C_{60}$-based devices, the decomposition began from the interface adjacent to HTL near Au metal electrode opposite to the case of $TiO_2$ based devices. Since the two types of devices have identical structure except for the ETL, $C_{60}$ or $TiO_2$/Perovskite/Spiro-MeOTAD/Au, these different degradation characteristics indicate that charge extraction may play an important role where moisture driven decomposition of perovskite material begins. Second, the $TiO_2$ based devices suffer from severe hysteresis, whereas the $C_{60}$ based devices do not. Considering that the origin of hysteresis is known as capacitive current[24], trapped charge[25] and unbalanced charge injection[26], many electrons may be accumulated near the ETL in the $TiO_2$ based devices, while the $C_{60}$ based devices hardly do. From the observation on the degradation of the $TiO_2$ based devices that begins from the interface contacting $TiO_2$ layer where many charges could be trapped, it is reasonable to suspect that trapped charges at the interface between perovskite and charge extraction layer would be responsible for initiating the moisture related decomposition. Fast extraction of electrons through $C_{60}$ would hardly accumulate negative charges at the interface between perovskite and $C_{60}$, but hole extraction through Spiro-MeOTAD could be slower than the rate of electron extraction in the $C_{60}$ based cell[27,28]. This could result in positive charge trapping at the interface between perovskite and hole extraction layer, which could be the cause why the degradation begins from the interface between perovskite and Spiro-MeOTAD for $C_{60}$ based cells (see Fig. 1e). These results demonstrating the degradation beginning from opposite side for different charge extraction layers gave us a clue about the trapped charge driven degradation regardless of polarity.

Another intriguing experimental observation is the light soaking in the presence of moisture which consistently showed irreversible degradation of perovskite in previous works[9,11,12] while in the dark condition moisture introduction only formed reversible hydrates of perovskites, for example, $CH_3NH_3PbI_3 \cdot H_2O$ or $(CH_3NH_3)_4PbI_6 \cdot 2H_2O$[11,15]. The reason has not been elucidated yet although Christians et al.[11] suggested that organic cation could become less tightly bound to $PbI_6^{4-}$ octahedra after light soaking. In the present study, along with the scenario of the above mentioned trapped charges that could trigger irreversible degradation, the charge generation under light soaking and subsequent trapping on the surface of perovskite is suspected to initiate the moisture induced irreversible degradation under light illumination. To confirm the irreversible degradation under light soaking, we also investigated the experiments under light soaking or not in the presence of moisture. Figure 2a, b showed the degradation behavior of MAPbI$_3$ and our mixed MA$_{0.6}$FA$_{0.4}$PbI$_{2.9}$Br$_{0.1}$, respectively, for two days in the dark condition with relative humidity (RH) 90 %. Absorption spectra measurements show that the original MAPbI$_3$ (black curve) became hydrated (red curve) after two days and then dehydrated reversibly via N$_2$ drying, which is consistent with previous studies[11,15]. On the other hand, the absorption spectra of our mixed MA$_{0.6}$FA$_{0.4}$PbI$_{2.9}$Br$_{0.1}$ perovskite were hardly changed with the same condition (see Fig. 2b) and XRD patterns were the same after two days (see Supplementary Fig. 6). This indicates our mixed composition perovskite would be more resistible to become hydrated than distorted tetragonal perovskite MAPbI$_3$. It is likely that water molecules could penetrate more easily into the distorted tetragonal MAPbI$_3$ than into the more compact cubic crystal structure of MA$_{0.6}$FA$_{0.4}$PbI$_{2.9}$Br$_{0.1}$. A slight change of the absorption spectra shown in Fig.2b indicates a slow hydration could still happen to our mixed perovskite under 90 % RH. However, both perovskites showed irreversible degradation rapidly under light soaking even at low RH 20 % (see Fig. 2c, d), which is also consistent with previous report[11,12,27]. It is interesting to note that under light soaking, our mixed perovskite degrades more slowly than the conventional MAPbI$_3$. As mentioned earlier, perovskite absorbing light can generate and store charges due to its capacitive property[29] that may be trapped on the grain boundaries (we will show trapped charges along grain boundaries after light soaking later). It could be hypothesized that these trapped charges generated under light soaking would be responsible for irreversible degradation, which is in line with the aforementioned hypothesis of trapped charge driven degradation explaining the initiation of degradation on different side depending on different charge extraction layers.

To prove this compelling hypothesis of trapped charge driven degradation, we have configured a novel experimental setup employing an ion generator by corona discharge and a stainless chamber that blocks all incident light from outside as shown in Fig. 3a. The air inside the chamber is isolated from the outside and controlled by two gas inlets that are connected to the independent gas sources (Gas 1 and Gas 2) (see more details in Supplementary Information: Experimental setup for ion generation and deposition). Gas 1 is ionized by applying a high voltage to the pin of the corona chamber, delivered to the deposition chamber by gas flow, and deposited on the perovskite film placed at the bottom of the deposition chamber. Gas 2 passes through a water bubbler to regulate the humidity level in the deposition chamber. We measured the time evolution of the absorption spectra as the perovskite films were gradually degraded in the deposition chamber. Gas 1 was chosen as nitrogen or hydrogen for generating positive or negative ions for being used as different polarity charges trapped on the surface of perovskite, respectively while Gas 2 was nitrogen (see more details in Supplementary Information: Corona ion generation). First, we needed to check that $N_2$ positive ions and $H_2$ negative ions themselves do not affect the degradation without moisture (see Supplementary Fig. 7). Next, we examined the degradation behavior in the presence of moisture when charged ions deposited on the surface of perovskite. When the positively charged $N_2$ ions were deposited and the relative humidity in the chamber was held at 40%, the perovskite film rapidly decays as shown in Fig. 3b1. The deposition of negatively charged $H_2$ ions also showed the similar irreversible degradation behavior under the same moisture level in Fig. 3b2. Although Fig.3b2 for negative charges appears to cause slower degradation than the case shown in Fig.3b1 for positive charges, this could not tell which polarity charges affect more adversely on the degradation since ion generation for different polarity in our experiment is different (see more details in Supplementary Information : Corona ion generation). Note that, on the other hand, without depositing charges, the absorption spectra and XRD patterns of the perovskite film were hardly changed for 2 days even under 90% humidity as was shown in Fig. 2b and Supplementary Fig. 6. Similarly, at the presence of only charges without moisture, the degradation did not occur at all as shown in Supplementary Fig. 7. This suggests that the irreversible degradation of perovskite materials only take place when both moisture and charges exist simultaneously. Moreover, the structurally distorted (conventional) $MAPbI_3$ film was degraded more quickly than the mixed stable $MA_{0.6}FA_{0.4}PbI_{2.9}Br_{0.1}$ under the same moisture and ion deposition level (Supplementary fig. 8). Based on these observations, the degradation mechanism could be thought of two-step process: the formation of hydrated perovskite by humidity and the irreversible decomposition by trapped charges. The first step of the formation of hydrated perovskite was already reported by several groups[10,11]. Here, we suggest that local electric field caused by trapped charge could distort electrostatically the structure of hydrated perovskite in which octahedral $PbX_6^{4-}$ interacts with both organic cation and $H_2O$ and trigger the initiation of irreversible decomposition of perovskite (will discuss more complete scenario later). This local electrostatic distortion seems to be different from macroscopic electric field that was previously reported as a possible cause on the irreversible degradation through ion movement[9,15]. We also checked this possibility by applying electric field of 600 V/cm under 90% RH without touching the perovskite surface to prevent the possibility of trapped charge generation between the electrode and perovskite material. Leijten et al.[9] found the irreversible degradation near the Au electrodes touching the perovskite by applying even a weak field of 600 V/cm in the presence of moisture. For our case, the degradation did not occur even under 90 % RH with 600 V/cm unless ion deposition is treated (Supplementary Fig. 9). For their case, it is possible that charges might have been trapped at the interface between Au electrode and perovskite, and then, in the presence of moisture, trapped charge driven irreversible decomposition could have begun from the interface under the electrode, which was also observed in their experiment. Note that electric field applied by two electrodes in their experiment would be rather spatially uniform on the perovskite between two electrodes, therefore, there was no reason that the degradation should have started from the electrode if electric field alone could cause irreversible degradation. It is likely that ion movement due to electric field could be possible only after perovskite is decomposed into ions such as MA+ and then externally given electric field could accelerate the degradation process.

Next we investigated how trapped charge could decompose perovskite material in time by examining morphology evolution via SEM analysis. As shown in Fig. 3c, the degradation is initiated from the grain boundaries. As the reaction continues, the color of the film turns into yellow, indicating that the perovskite is irreversibly decomposed to $PbI_2$ (see XRD patterns after degradation in Supplementary Figure 10). It is interesting to dig into why degradation occurs from grain boundaries in line with our trapped charge mechanism. To check the distribution of trapped charges on the surface of perovskite after uniform ion deposition, we measured Kelvin Probe Force Microscopy (KPFM) of un-treated perovskite and ion-treated perovskite films. Fig. 3d shows topology and surface potential distribution of the perovskite surface on which positive $N_2$ ions were uniformly showered. Striking coincidence between two images is the evidence that charges are preferentially trapped along grain boundaries. Overlapped image of topology and potential distribution shown in Supplementary Fig. 11 clearly demonstrates charges are trapped along grain boundaries even though ions are showered uniformly. For untreated sample, there is no correlation between topology and potential distribution (see Supplementary Figure 11). With this charge trapping along grain boundaries, experimentally observed degradation pattern following grain boundaries and the fact that the irreversible degradation occurs only when moisture and charges exist together are the evidences that trapped charges would be responsible for the initiation of irreversible degradation of perovskite materials. It is now apparent that grain boundaries are the most vulnerable sites for the degradation because they provide charge accumulation sites as well as infiltration pathway of water vapor[9]. Successful enhancement of stability utilizing high mobility inorganic charge extraction layers supports the present idea [13,18].

To further investigate the possibility that the above mentioned intriguing experimental observation of irreversible degradation under light soaking might be related to the mechanism of trapped charges of the present study, we measured KPFM images on the surface of perovskite after light illumination without ion deposition. As shown in Supplementary Fig. 11, charges are clearly trapped along grain boundaries for the sample soaked by light confirming that light soaking alone induces charge trapping along the grain boundaries of perovskite material like was done by introduction of ion charges in the dark condition. From our concept, these trapped charges can now trigger the irreversible degradation due to moisture as the same happened when ion charges are deposited in the dark. Therefore, the fundamental cause for irreversible degradation would be the same, that is, the trapped charges that could trigger the irreversible degradation under humid air. Such irreversible degradation under light soaking was reported in previous several reports[11,20], but, the reason has not been clearly elucidated so far although Christians et al.[11] suggested the lessened hydrogen bonding after photoexcitation as a possible cause. Here, we argue that trapped charges under moisture would be responsible for the initiation of irreversible degradation under light or in the dark since light illumination always generates charges and traps the charges along grain boundaries as shown earlier. This can explain well why moisture itself without illumination or intentional ion deposition only hydrated perovskite reversibly. Light illumination under nitrogen gas without moisture for two days was shown to hardly degrade the perovskite as shown in Supplementary Fig. 12 which is strongly contrasted with the case of light illumination under moisture (see Fig. 2d).

We suggest a scenario how trapped charge could trigger the irreversible decomposition of perovskite materials. First, in the presence of water molecules, perovskite materials are known to form hydrates. Within the hydrated perovskite, octahedral $PbX_6^{4-}$ interacts with both organic cations ($CH_3NH_3^+$, $HC(NH_2)_2^+$) and $H_2O$[11]. Then, the charges trapped at the defect site regardless of polarity could help to deprotonate organic cations by induced local electric field like the way that was well known from the soft matter physics on electric-field induced de-protonation of organic molecules[31,32]. Such de-protonation process in the presence of water yields volatile molecules like $CH_3NH_2$ and $HC(=NH)NH_2$ that can evaporate at room temperature. The following deprotonation from organic cations could take place due to trapped charge induced local electric field:

$$\begin{pmatrix} CH_3NH_3^+\,(MA^+) \\ HC(NH_2)_2^+\,(FA^+) \end{pmatrix} + H_2O \xrightarrow{TC} \begin{pmatrix} CH_3NH_2(\uparrow) \\ HC(=NH)NH_2(\uparrow) \end{pmatrix} + H_3O^+ \quad \ldots (1)$$

where TC means trapped charge. Evaporation of resulted volatile neutral molecules could shift the following equilibrium reaction that prevail during the formation of hydrates towards the right hand side, which causes the beginning of irreversible degradation of perovskite:

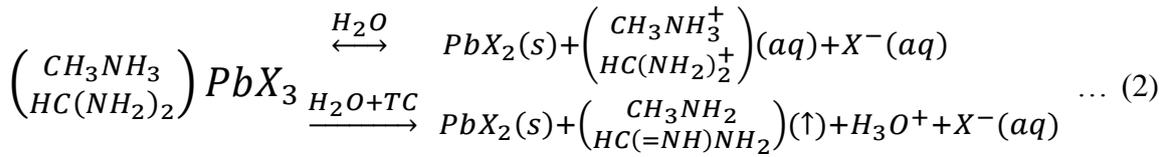

$$\begin{pmatrix} CH_3NH_3 \\ HC(NH_2)_2 \end{pmatrix} PbX_3 \underset{H_2O+TC}{\overset{H_2O}{\rightleftharpoons}} \begin{matrix} PbX_2(s)+\begin{pmatrix} CH_3NH_3^+ \\ HC(NH_2)_2^+ \end{pmatrix}(aq)+X^-(aq) \\ PbX_2(s)+\begin{pmatrix} CH_3NH_2 \\ HC(=NH)NH_2 \end{pmatrix}(\uparrow)+H_3O^++X^-(aq) \end{matrix} \quad \ldots (2)$$

where X denotes halide. In addition, trapped charges can help the hydration process by distorting the structure of perovskite electrostatically, which leads water molecule to penetrate easily into the perovskite structure.

Next, we investigated the effect of oxygen on the moisture induced degradation of perovskite materials. Previous studies on device encapsulation have mostly focused on blocking moisture, and overlooked the effect of oxygen. However, we found a remarkable result that is the synergetic effect of oxygen on the process of moisture induced degradation of perovskites. To test this, we switched the Gas 2 from $N_2$ to dry air ($N_2+O_2$:8:2) for bubbling water (see Fig. 3a) with the Gas 1 remaining as $N_2$ for ion generation. As shown in Fig. 4b, the addition of oxygen under the same condition of trapped charge and moisture of RH=40% clearly showed more rapid degradation compared to the case without oxygen (Fig. 4a). We also verified that degradation did not happen if a dry air gas was employed as Gas 2 without moisture (Supplementary Fig. 12). This implies that oxygen alone would not harm the perovskite even under the existence of trapped charges, but, the oxygen could worsen the degradation process in the presence of water and trapped charges. Previously, Niu et al.[30] suggested reaction equations of oxygen-involved degradation in humid air condition, which represent the formation of H₂O as a reaction product. Similarly, we explain the fast degradation in the presence of oxygen more clearly based on the scavenging effect of oxygen on H₃O+ proton generated from irreversible de-protonation process, Eq(1). The process of scavenging $H_3O^+$ by oxygen could be expressed as follows.

$$H_3O^+ + X^- + \frac{1}{4}O_2 \rightarrow \frac{1}{2}X_2 + \frac{3}{2}H_2O \dots (3)$$

The overall chemical reaction of the oxygen involved degradation can be expressed as:

$$\begin{pmatrix} CH_3NH_3 \\ HC(NH_2)_2 \end{pmatrix} PbX_3(s) + \frac{1}{4}O_2 \xrightarrow{H_2O+TC} PbX_2(s) + \begin{pmatrix} CH_3NH_2 \\ HC(=NH)NH_2 \end{pmatrix}(\uparrow) + \frac{1}{2}X_2 + \frac{1}{2}H_2O \dots (4)$$

Interestingly, the overall reaction produces water, which is in agreement with the previous work[30], and then this water causes a chain reaction of water induced degradation. Therefore, oxygen, which comprises about 20% of the atmosphere, should be considered as an additional target that must be avoided together with moisture.

In conclusion, we found that trapped charges would be responsible for triggering the irreversible degradation in the moisture induced degradation of perovskite materials regardless of charge polarity. To verify this, we designed a novel experimental setup enabling the deposition of different polarity charges on the surface of perovskites for controlled moisture degradation experiments. From this setup, we demonstrated that the perovskite materials degraded irreversibly along grain boundaries only when both moisture and trapped charge exist simultaneously. During the course of study, we developed and used a new mixed perovskite material that ensured both high performance and structural stability. Our study explains both why the degradation begins to occur from the different side of interface between perovskite and charge extraction layer for different charge extraction layers and how light soaking always degrades irreversibly in the presence of moisture. KPFM study reveals that charges are trapped preferentially along grain boundaries of perovskites even under uniform deposition of ion charges or uniform illumination of light, which supports our idea of trapped charge driven degradation. We also found the synergetic effect of oxygen in the process of moisture induced degradation. The present study suggests that the prevention of accumulation of charges at the interface is very important in addition to proper encapsulation for developing commercially viable perovskite solar cells.

**Acknowledgements**

This work was supported by the Global Frontier R&D Program of the Center for Multiscale Energy Systems funded by the National Research Foundation under the Ministry of Education, Science and Technology, Korea (2011-0031561 and 2011-0031577). We thank H. Sung and S.R. Noh for discussions and comments on this work.

**Author Contributions**

N.A., K.K. and M.C. conceived and designed the experiments and analyzed the data results. N.A., K.K., H.Y., and J.B. performed the device fabrication and photovoltaic performance measurements. N.A., K.K., and J.B. carried out the controlled stability test. B.Y.L. and J.-K.L. measured Kelvin Probe Force Microscopy. P.V.P., N.A., K.K., and M.C. discussed the mechanism. M.C. led the project. N.A., K.K., M.S.J, P.V.P., and M.C. wrote the paper. All authors discussed the results.

Fig.1

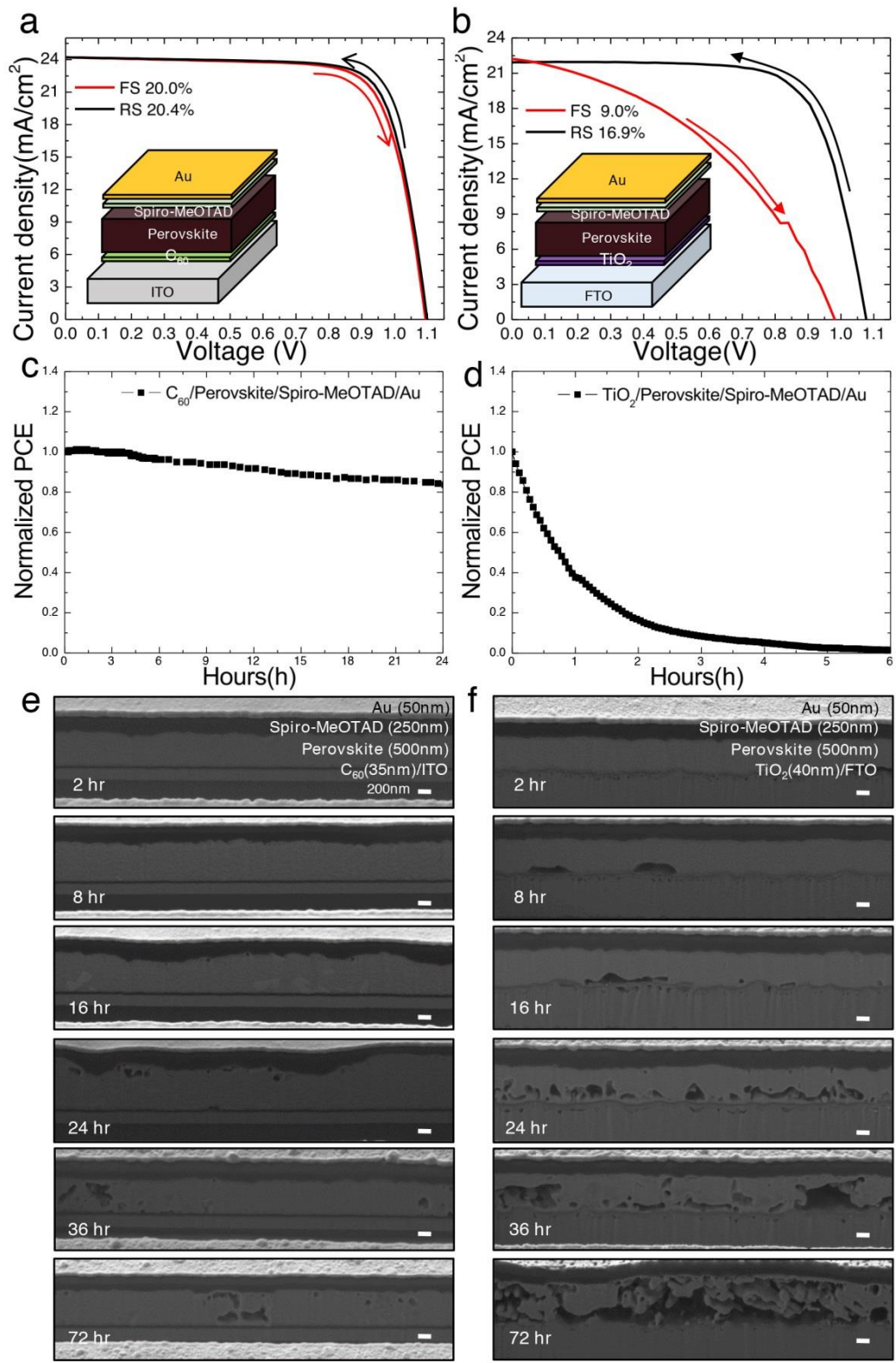

**Figure 1.** J–V curves of (a) ITO/C$_{60}$(35nm)/MA$_{0.6}$FA$_{0.4}$PbI$_{2.9}$Br$_{0.1}$(500nm)/Spiro-MeOTAD(250nm)/Au(50nm) and (b) FTO/TiO$_2$(40nm)/MA$_{0.6}$FA$_{0.4}$PbI$_{2.9}$Br$_{0.1}$(500nm)/Spiro-MeOTAD(250nm)/Au(50nm) measured in the reverse (black) and forward (red) scans with a 200 ms sweep delay. (c, d) Time evolution of the normalized PCE measured under one sun illumination in ambient conditions (relative humidity = 30%) of the (c) C$_{60}$ and (d) TiO$_2$ based devices. (e, f) Time evolution of the FIB-SEM cross-sectional images of the (e) C$_{60}$ and (f) TiO$_2$ based devices aged for 72 h under light illumination in ambient conditions. Scale bars = 200 nm.

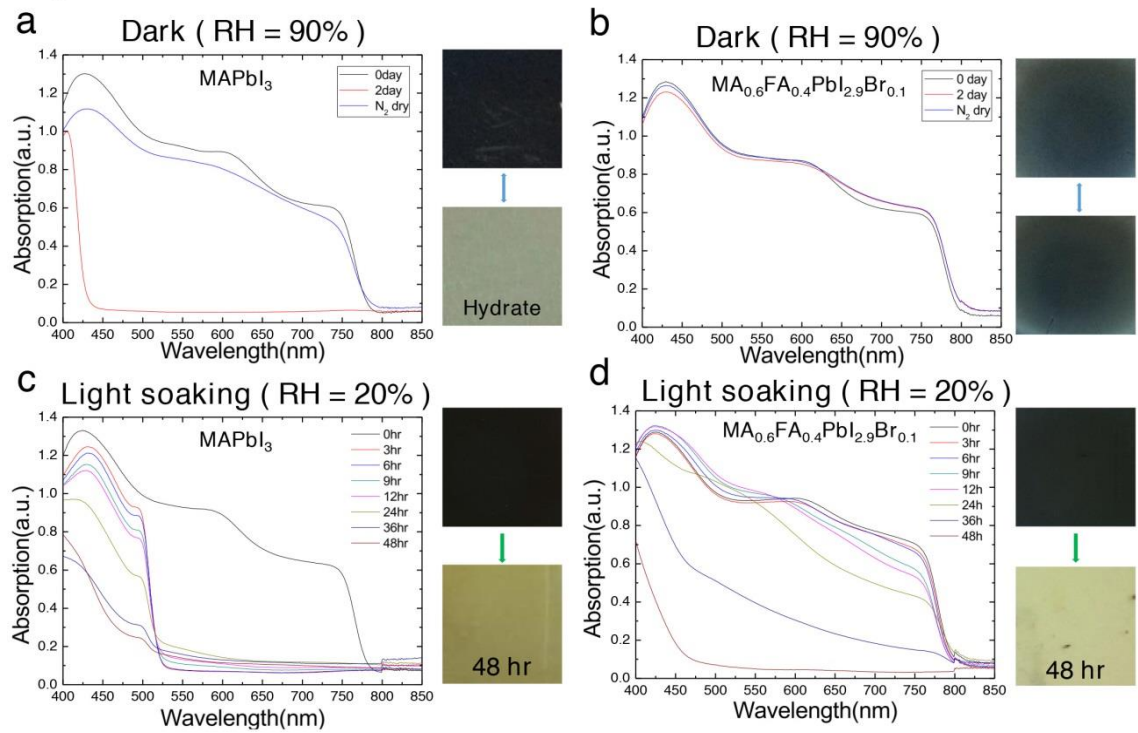

**Figure 2.** Absorption spectra of (a) MAPbI$_3$ and (b) MA$_{0.6}$FA$_{0.4}$PbI$_{2.9}$Br$_{0.1}$ perovskite films under dark conditions at 90% relative humidity. MAPbI$_3$ perovskites were transformed into transparent hydrated states after 2 days. Time evolution of absorption spectra of (c) MAPbI$_3$ and (d) MA$_{0.6}$FA$_{0.4}$PbI$_{2.9}$Br$_{0.1}$ during light soaking at 20% relative humidity. Pictures of the perovskite films before and after aging are shown on the left side of each figure.

Fig.3

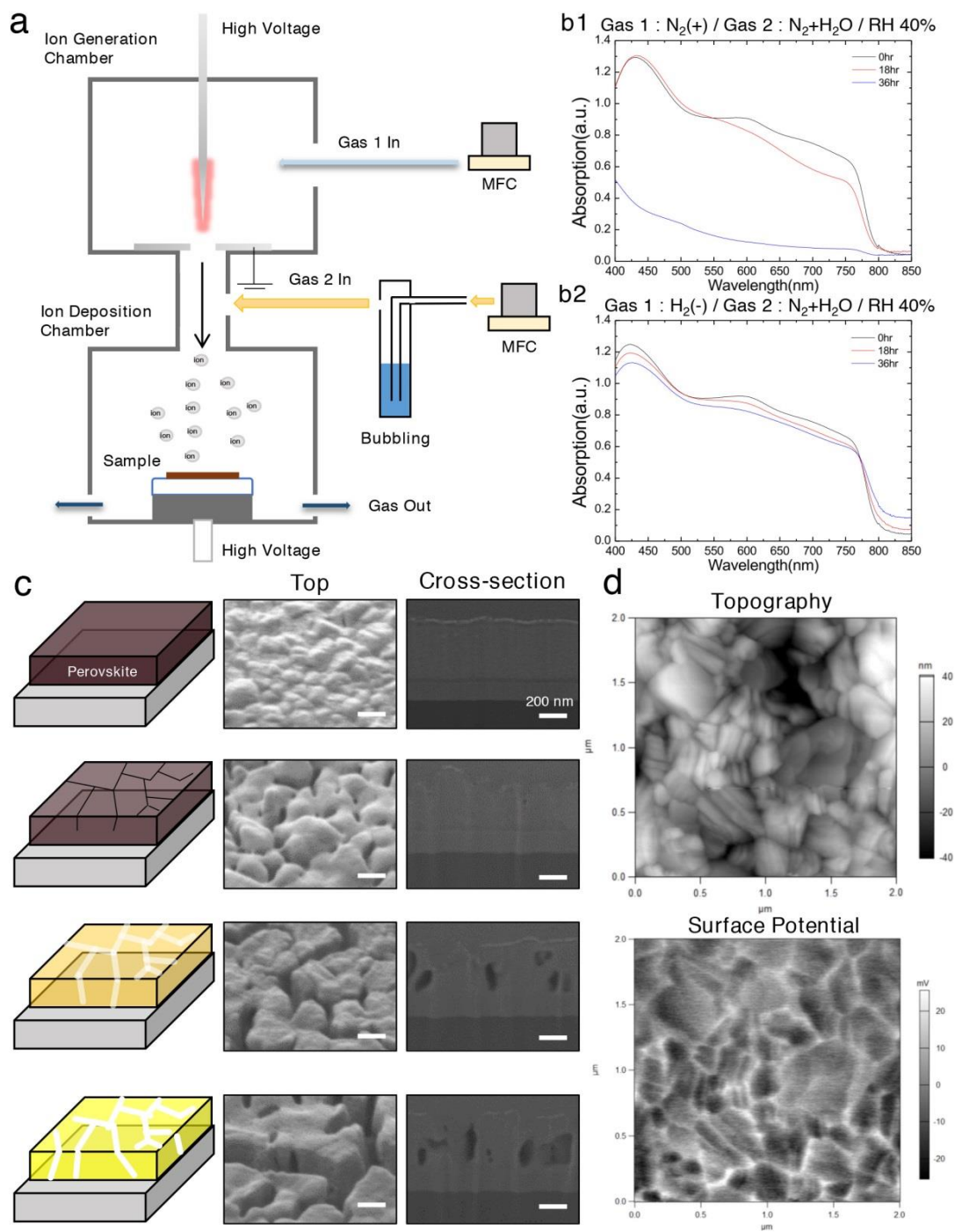

**Figure 3.** (a) Experimental setup of corona discharge for ion generation, bubbling system for humidification, and SUS chamber for ion deposition and blocking light. (b) Absorption spectra of the perovskite film measured at an interval of 18 hours during deposition of (b1) positive nitrogen ions and (b2) negative hydrogen ions at 40% relative humidity. (c) Scheme description of perovskite degradation processes (left), and top-view (middle) and cross-sectional (right) SEM images of perovskite layers by ion deposition in humidified nitrogen. Scale bars = 200 nm. (d) Topography and surface potential profile of $MA_{0.6}FA_{0.4}PbI_{2.9}Br_{0.1}$ film obtained from KPFM measurements after deposition of $N_2$ positive ions.

Fig. 4

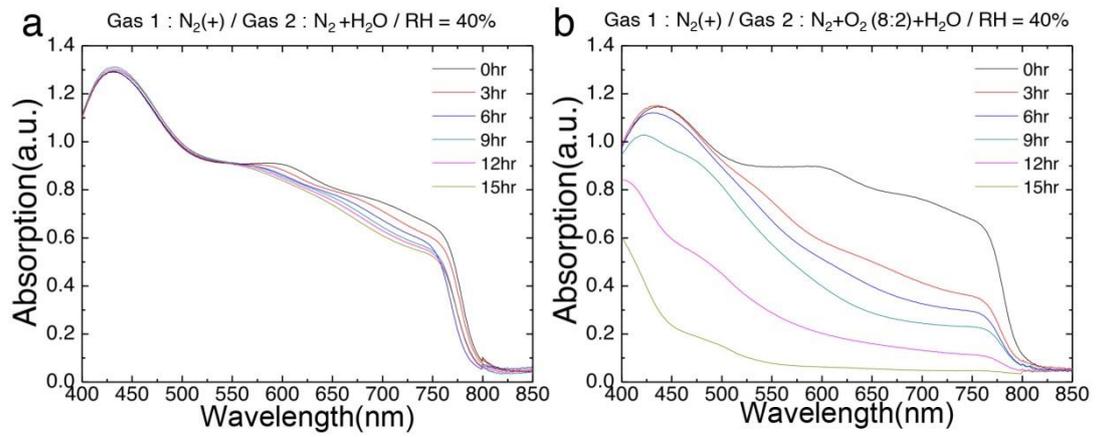

**Figure 4.** Time evolution of the absorption spectra of $MA_{0.6}FA_{0.4}PbI_{2.9}Br_{0.1}$ films (a) in humidified nitrogen and (b) in humidified air measured at an interval of 3 hours during deposition of positive nitrogen ions. In both cases, the relative humidity is 40%.

# Supplementary Information

# Trapped charge driven degradation of perovskite solar cells

Namyoung Ahn[1,2,†], Kwisung Kwak[1,2,†], Min Seok Jang[1], Heetae Yoon[1,2], Byung Yang Lee[3], Jong-Kwon Lee[1], Peter V. Pikhitsa[1], Junseop Byun[1,2], and Mansoo Choi[1,2,*]

[1] Global Frontier Center for Multiscale Energy Systems, Seoul National University, Seoul 151-742, Korea

[2] Department of Mechanical and Aerospace Engineering, Seoul National University, Seoul 151-742, Korea

[3] Department of Mechanical Engineering, Korea University, Seoul 136-713, Korea

[†] These authors contributed equally to this work.
*To whom correspondence should be addressed. E-mail: mchoi@snu.ac.kr

## Optimization of mixed MA$_x$FA$_{1-x}$PbI$_y$Br$_{3-y}$

In a mixed cation and halide anion system of Pb-based perovskite, APbX$_3$, the Goldschmidt tolerance factor t is defined in terms of the average radii of ions as follows.

$$t = \frac{\overline{r_A}+\overline{r_x}}{\sqrt{2}(\overline{r_{Pb}}+\overline{r_x})} \quad \dots (1)$$

where $\overline{r_A}$, $\overline{r_X}$, and $\overline{r_{Pb}}$ are the average radii of the cation, anion, and lead ions, respectively. The I$^-$, Br$^-$ and Pb$^{2+}$ ions have a spherical shape and the radii of these ions have been measured in previous works[1]. In constrast, it is difficult to precisely estimate the radii of the organic cations perched on the A site due to their non-spherical shape and rotational motion, nevertheless, several groups already reported the calculated effective ion radius of both MA+ and FA+ cation[2,3,4,5]. By using equation 1 and the suggested ionic radii ($r_{MA+}$= 0.18 nm, $r_{I-}$= 0.22 nm, $r_{Pb2+}$= 0.12nm), the tolerance factor of MAPbI$_3$ is calculated to be 0.83[5], indicating that MA+ cations are too small to fit into the interstices between PbX$_6$ octahedra. This mismatch causes crystal distortion, and consequentially MAPbI$_3$ perovskite have unstable tetragonal crystal structure[6,7]. By partially replacing MA+ and I- to relatively larger FA+ and smaller Br- ions respectively, crystal distortion can be alleviated to produce more stable cubic structure with the tolerance factor between 0.9-1[8]. Simple calculation reveals that replacing I- to Br-($r_{Br-}$= 0.196 nm) only marginally affects the perovskite crystal structure as the tolerance factor of MAPbBr$_3$ is only 0.01 higher than that of MAPbI$_3$. On the other hand, replacing MA to FA cation can significantly alter the crystal structure because FA cations are expected to be much larger than MA cations. The exact radius of FA cation is still controversial[8], but considering that FAPbI$_3$ can possess not only non-perovskite yellow δ-phase with hexagonal crystal structure (t >1) but also black perovskite α-phase with cubic structure (0.9 < t < 1), it can be speculated that the tolerance factor of $FAPbI_3$ would be around 1. The radius of FA cation $r_{FA+}$ is estimated to be around 0.26 nm from these speculations and equation 1. The tolerance factor of mixed cation system, $MA_xFA_{1-x}PbI_3$, is calculated as a function of the ratio x as shown in Fig. S1. This relation suggests that the x values between 0.2 and 0.6 lead to the most stable cubic crystal structure, as the resulting tolerence factor lies between 0.9 and 1.

We measured X-ray diffraction(XRD) patterns of five types of mixed perovskite films (x= 1, 0.8, 0.6, 0.4, 0.2) coated on the ITO glass in order to compare their stability. Fig. S2a,b show the XRD patterns of the fresh and degradaded samples, respectively. The perovskite films were degraded in the chamber (relative humidity (RH) ~ 50%) for 10 hours under one sun illumination. The peak originating from $PbI_2$ at 12.7 degrees intensively appears in the case of $MAPbI_3$ (x=1), and the peak of non-perovskite δ-phase $FAPbI_3$ at 11.7 degrees appears for x < 0.4, which indicates severe instability to water vapor[10,11]. (Supplementary Fig. 2c,d) These observations suggest that the most stable composition would be $MA_{0.6}FA_{0.4}PbI_3$ (x=0.6), which still posesses the tolerance factor ensuring the cubic crystal structure.

We also fabricated the full devices with the ITO/$C_{60}$/Perovskite/Spiro-MeOTAD/Au structure in order to find the optimal composition with respect to the photovoltaic performances in mixed perovskite $MA_xFA_{1-x}PbI_yBr_{3-y}$. Supplementary Fig. 3a,b summarize the PCEs as a function of the MA+ fraction(x) and the Br- fraction(3-y), respectively. As a result, the $MA_{0.6}FA_{0.4}PbI_{2.9}Br_{0.1}$–based devices were shown to the best performance. Considering that MA:FA=0.6:0.4 is the best composition for stability, we concluded that $MA_{0.6}FA_{0.4}PbI_{2.9}Br_{0.1}$ would be the best composition in terms of both performance and stability in mixed perovskite systems.

**Solar cell fabrication**

Indium tin oxide-coated (ITO) glass substrates (AMG, 9.5Ω/cm$^2$, 25×25 mm$^2$) were sequentially sonicated in acetone, isopropanol, and deionized water. The cleaned substrate was sufficiently dried in oven in order to eliminate all residual solvents. A 35nm thick $C_{60}$ layer was densely coated on the ITO glass substrates by using a vacuum thermal evaporator at the constant rate of 0.1Å/s. The 50 wt% mixed perovskite solutions (MAI+FAI+MABr : $PbI_2$ : DMSO = 1:1:1 in DMF solvent) were coated on the ITO/C60 substrate by Lewis base adduct method[11]. To prepare our best compositional solution, 461 mg of $PbI_2$, 79.5 mg of MAI, 68.8 mg of FAI, 11.2 mg of MABr, and 78 mg of DMSO were mixed in 0.55 ml of DMF at room temperature with stirring for 30 min. After spin coating at 4000 rpm for 20s with ether dripping treatment, the transparent adduct films were annealed at 130 °C for 20 min to form black perovskite films. To prepare the hole trasport material(HTM), 72.3mg of Spiro-MeOTAD(Merk) dissolved in 1mL Chlorobenzen(Sigma-Aldrich) with 28.8 μl of 4-tert-butyl pyridine and 17.5 μl of lithium bis (trifluoromethanesulfonyl) imide (Li-TFSI) solution (520 mg Li-TSFI in 1 ml acetonitrile (Sigma–Aldrich, 99.8%)). The HTM solutions were spin-coated onto the perovskite layer at 2000rpm for 30s. After all process, 50nm gold (Au) as a counter electrode was deposited on the HTM at the rate of 0.3 Å/s by using a vacuum thermal evaporator. Optimization study on $C_{60}$ based $MAPbI_3$ solar cells is currently in preparation[11].

In the case of $TiO_2$-based devices, a $TiO_2$ blocking layer was fabricated on FTO-coated glass substrates by spin-coating 0.15 M titanium di-isopropoxide dis(acetylacetonate) (Sigma-Aldrich, 75 wt% in isopropanol) in 1-butanol (Sigma-Aldrich, 99.8%) at sequentially increasing spin rate of 700 rpm for 8 s, 1000 rpm for 10 s and 2000 rpm for 40 s. After spin-coating, the $TiO_2$ blocking layer was heated at 125 °C for 5 min, and this process was repeated once again. The substrate was annealed at 550 °C for 1 hr. The rest of the processes are identical to the fabrication of $C_{60}$-based devices.

**Characterization**

The cross-sectional and surface images of the perovskite films and the fabricated perovskite solar cells were obtained from a high-resolution scanning electron microscope with a focused ion beam system (Carl Zeiss, AURIGA). The optical absorption spectra of the perovskite films coated on the ITO substrate were measured by UV-vis spectrophotometer (Agilent Technologies, Cary 5000) in the 400-850 nm wavelength range. The XRD patterns of the perovskite films on the ITO glass were collected by using New D8 Advanced (Bruker) in the 2θ range of 5-80 degrees. Photocurrent density-voltage curves were measured by a solar simulator (Oriel Sol3A) with keithley 2400 source meter under AM1.5G, which is calibrated to give 100 mW/cm2 using a standard Si photovoltaic cell (Rc-1000-TC-KG5-N, VLSI Standards). The J-V curves were measured by covering devices with a metal mask having an aperture. (6.76 $mm^2$) External quantum efficiency (EQE) was measured by a specially designed EQE system (PV measurement Inc.) with 75 W Xenon lamp (USHIO, Japan) as a source of monochromatic light. .

**Experimental setup for ion generation and deposition[13]**

The whole chamber contains two connected chambers: the ion generation(IG) chamber and the ion deposition(ID) chamber. The IG chamber has cylindrical shape with 30mm diameter and and 35 mm height. It is made of transparent acrylic which makes it possible to see the state of corona discharge during experiment. Stainless steel pin and plate creates highly asymmetric electric field in the chamber, when a bias voltage is applied between the pin and the plate. Gas 1 can flow into the chamber through an inlet on the side wall. The polarity of generated ions is determined by the polarity of applied voltage to the pin. The current of generated ions was measured by Faradaycup electrometer (Keithley Sub-femtoamp romote sourcemeter, 6430). A high-voltage supply (FuG ElektronikGmbH, HCP140-12500) apply voltage to the pin and the substrate in the ID chamber.

The ions generated in the IG chamber flows through a 115 mm long pipe with a 1.5mm diameter that connects the IG and ID chambers. A tee tube connected to this pipe introduces Gas 2 into the system. The flow rates of Gas 1 and 2 are both controlled by mass flow controllers (MKS instruments, MFC Controller 247D, MFC 1179A). The negatively (postively) charged gas ions are electrostatically attracted and deposited on the positively (negatively) biased substrate with the bias voltage of 2kV (-2kV).

**Corona ion generation**

Nitrogen gas inflow with the flow rate of 2 lpm was transformed into positive nitrogen ions by applying 4.2 kV to the pin. The electric current of generated nitrogen ions was measured to be 20.6-25.8 pA, which is indicative of positive ion generation. To generate negative ions, we used hydrogen gas with the flow rate of 2 lpm and applied negative bias (-1.55 kV) to the pin, whcich generates the current of -3.2 ~ -4.8 pA. (see **Table 1**)

**Atmosphere control**

To maintain the condition of the air inside the chamber at constant, the flow rate of Gas 2 was controlled by MFC and set as 1.5 lpm throughout the measurements. Nitrogen, dry air, humidified nitrogen and air were used as Gas 2. Nitrogen and hydrogen gases are highly purified by 99.999%, and dry air consists of 80% of nitrogen (99.999%) and 20% of oxygen(99.995%). Gas 2 passes through a water bubbler that controls the humidity in the chamber. The relative humidity was measured by portable multifunction data-logger(Delta OHM, Data logger DO9847, Temp&Humidity probe HP474AC) at the gas exit of deposition chamber.

**Table 1** The parameters for ion generation and data obtained from MFC controller and Faradaycup

|      |          | Flow rate(L/m) | Current of generated ion (pA) | Applied voltage   |
|------|----------|----------------|-------------------------------|-------------------|
| Gas1 | Nitrogen | 2              | 20.6 ~ 25.8                   | 4.2kV(0.009mA)    |
|      | Hydrogen | 2              | -3.2 ~ -4.8                   | -1.55kV(-0.4mA)   |
| Gas2 | Nitrogen | 1.5            |                               |                   |
|      | Dry air  | 1.5            | -                             |                   |

## Topography and Kelvin probe force microscopy

All the samples for topography and Kelvin probe force microscopy measurements were prepared on ITO glass substrates. The perovskite films were spin-coated on the ITO glass and ITO/$C_{60}$ substrate. The ion-treated sample was prepared by depositing $N_2$ positive corona ion for 1 hr on the ITO/perovskite substrate. In the case of the light-illuminated sample, the ITO/$C_{60}$ substrate was used in order to measure positive charge accumulation profile. After one sun illumination for 1 hr, the sample was measured under light on or off during KPFM operation(See Supplementary Fig. 9).

Topography and Kelvin probe force microscopy (KPFM) signals were measured by using an atomic force microscope (MFP-3D, Asylum Research, USA) with a Pt-coated tip with the spring constant of 2 nN/nm and the resonant frequency of 77 kHz. For each line scanning, topography was first measured and successively the surface potential was measured while scanning the same line at a fixed distance (30nm) above the sample surface. The surface potential was measured using an active electronic feedback circuitry: the bias voltage to the tip was modulated in order to equate the potential of the tip with that of the surface, resulting in minimum vibration amplitude of the AFM tip at the fundamental frequency. A 150W halogen bulb with single fiber light guide was used for illumination.

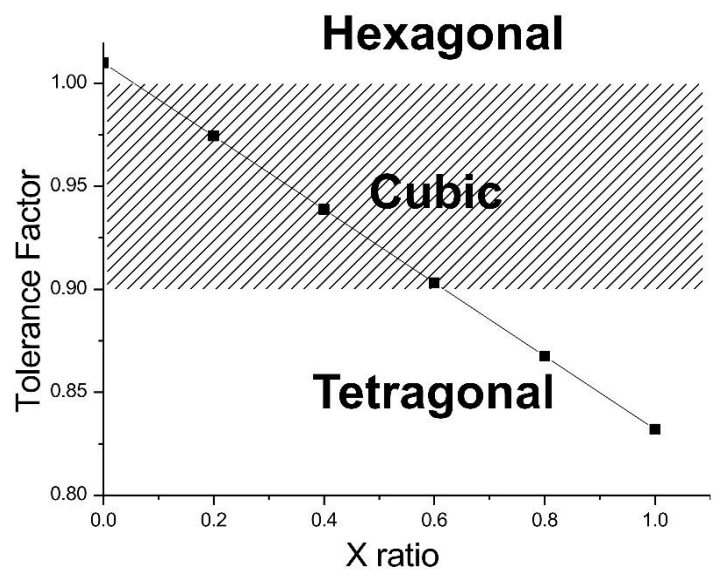

**Figure S1.** Dependence of the Goldschmidt tolerance factor on the MA fraction (x) of MAxFA1-xPbI3

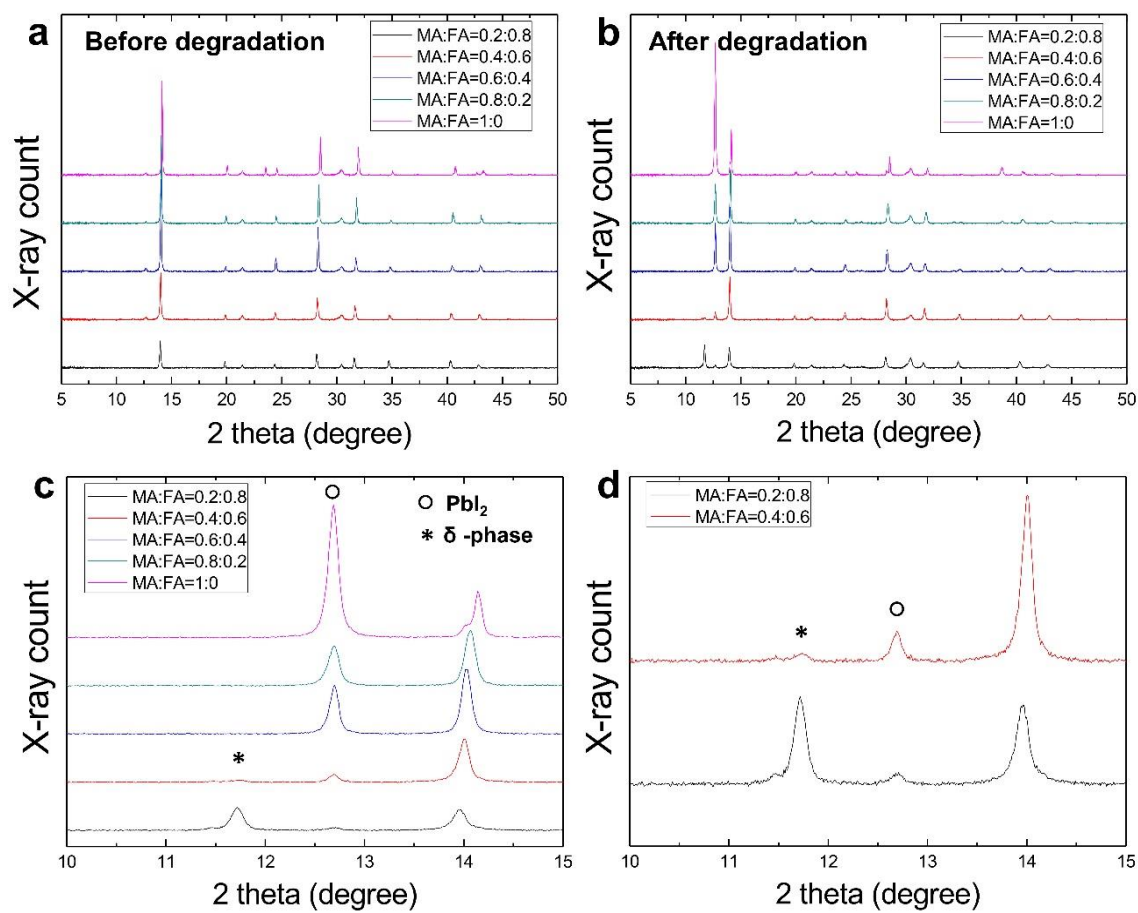

**Figure S2.** X-ray diffraction patterns of five different mixed perovskite films coated on ITO glass (x=0.2 (black), 0.4 (red), 0.6 (blue), 0.8 (green), and 1 (pink)) (a) before and (b) after degradation under one sun illumination at 50 % relative humidity for 10 hours. (c) Magnified XRD patterns around the peaks originating from $PbI_2$ and non-perovskite δ–phase. (d) Magnified XRD patterns for x=0.2 and 0.4.

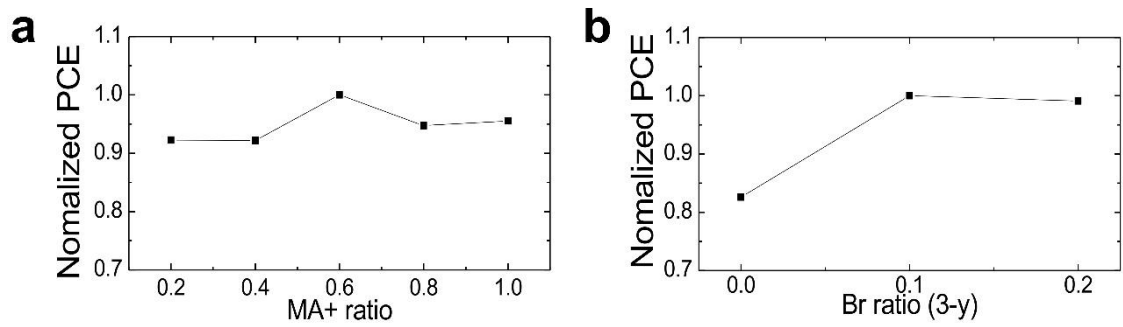

**Figure S3**. Dependence of the normalized PCEs of ITO/C$_{60}$/Perovskite/Spiro-MeOTAD /Au device on (a) the MA+ fraction (x), and (b) the Br- fraction (3-y)

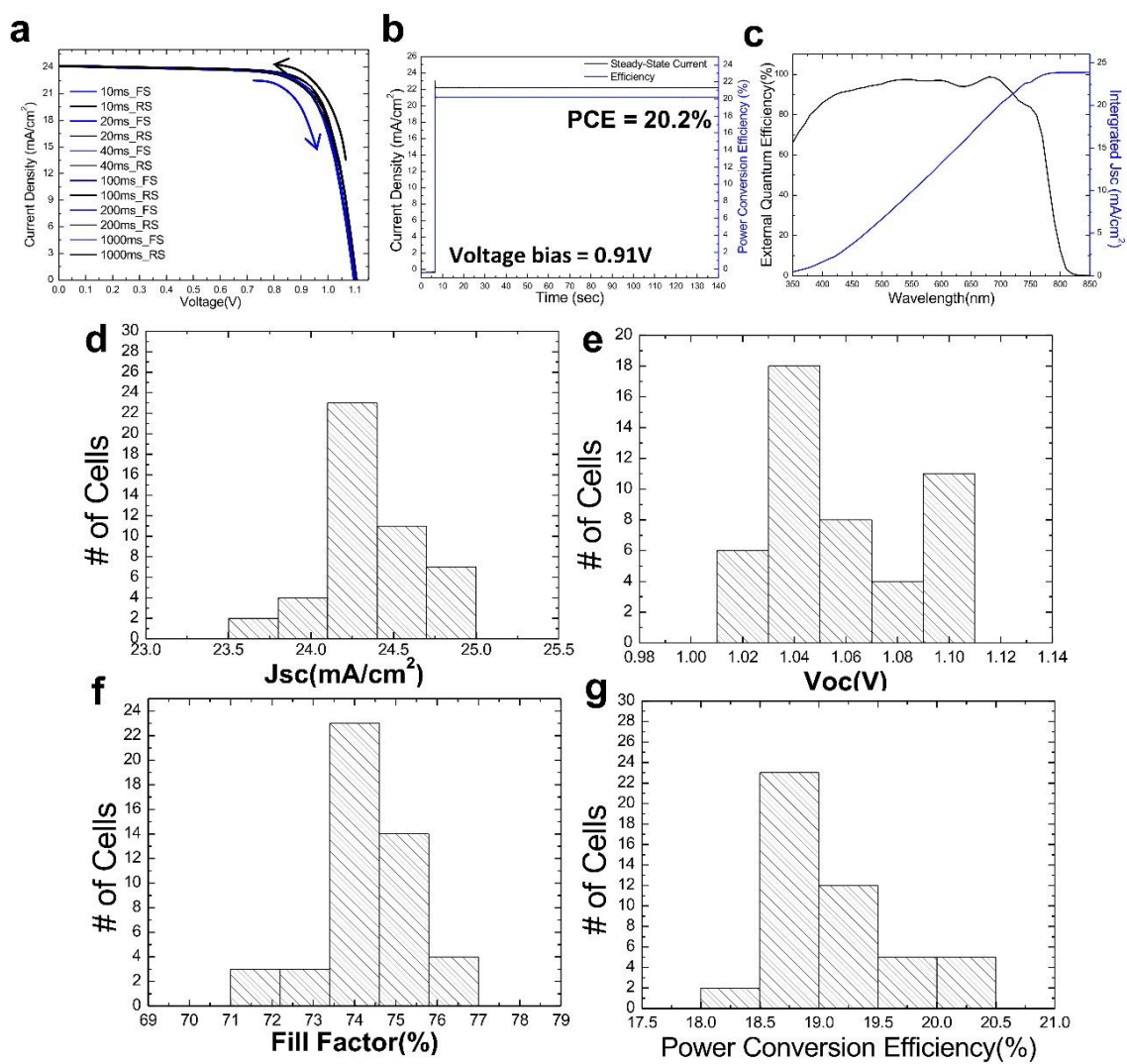

**Figure S4.** Photovoltaic performance characteristics. (a) J-V curves of the best-performing device measured at various sweep delay times. (b) Stabilized photocurrent density (black) and power conversion efficiency (blue) measured at a bias voltage of 0.91V for 140 seconds. (c) External quantum efficiency (EQE) spectrum and the integrated Jsc estimated from the measured EQE. Histograms of (d) short-circuit current density (Jsc), (e) open-circuit voltage (Voc), (f) fill factor (FF), and (g) power conversion efficiency (PCE) of 47 cells

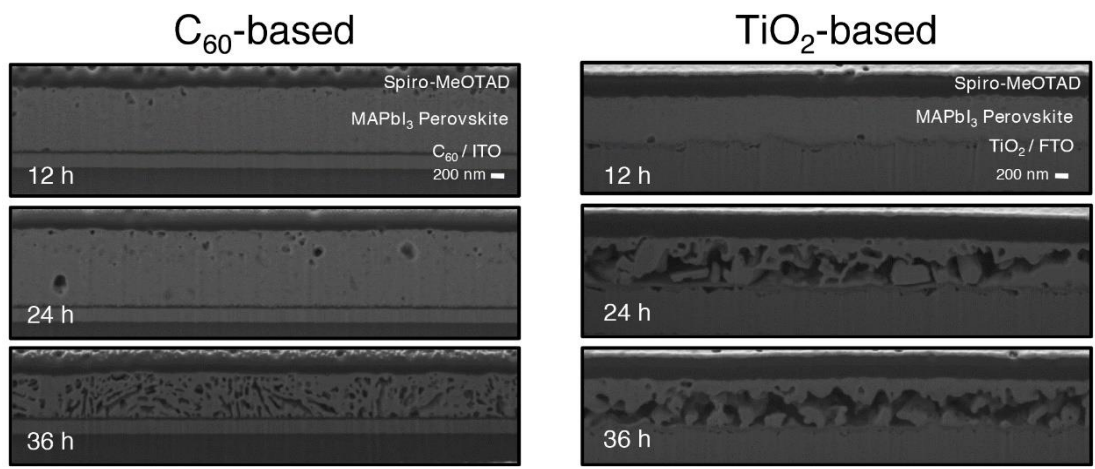

**Figure S5.** Degradation patterns of the $C_{60}$- (left) $TiO_2$- (right) based devices employing $MAPbI_3$ perovskite, which were aged for 36 h under one sun illumination in ambient conditions. Scale bars = 200 nm. As was shown in Fig. 1 for our mixed perovskite, the cell employing $MAPbI_3$ also showed the same pattern of degradation: for $TiO_2$-based cell, the degradation begins from the interface between $MAPbI_3$ and $TiO_2$ electron extraction layer, but for $C_{60}$-based cell, the degradation begins from the opposite side of interface between $MAPbI_3$ and Spiro-MeOTAD hole extraction layer.

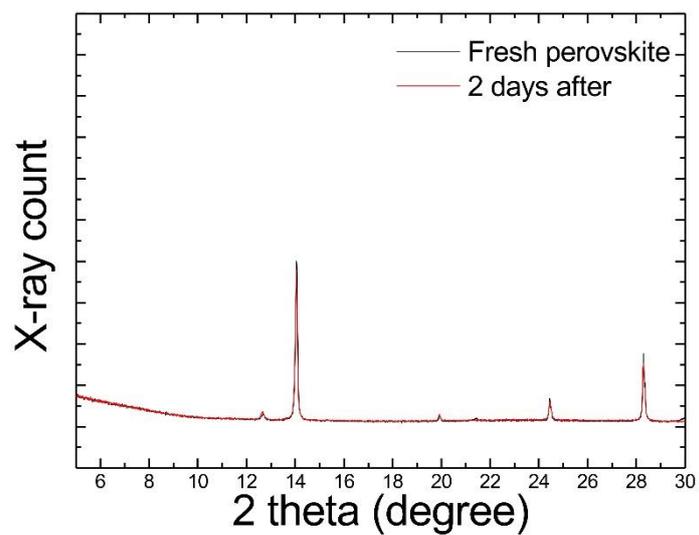

**Figure S6.** XRD patterns of the perovskite film before (red) and after (black) aged at 90% relative humidity for 2 days.

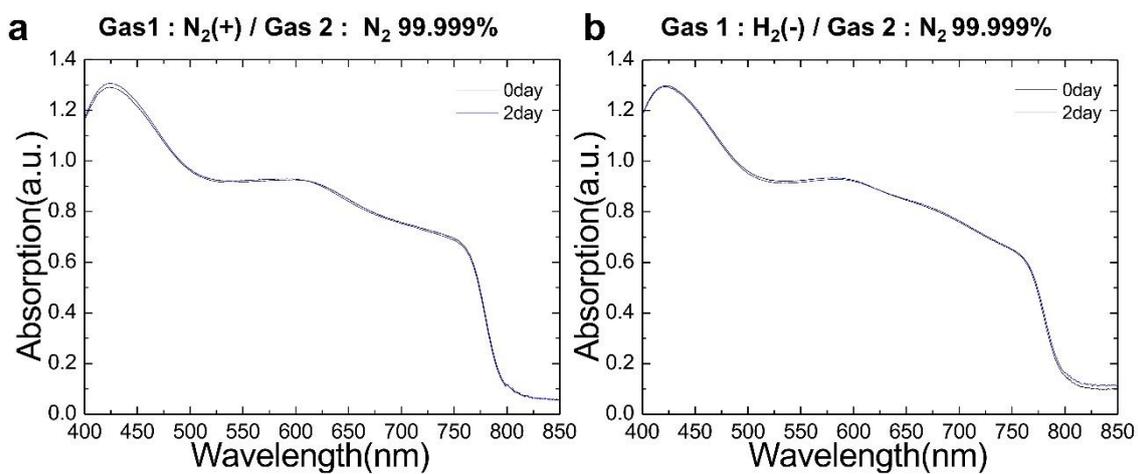

**Figure S7.** Absorption spectra of the perovskite film under (**a**) continuous positive nitrogen and (**b**) negative hydrogen ion deposition in moisture-free dark condition.

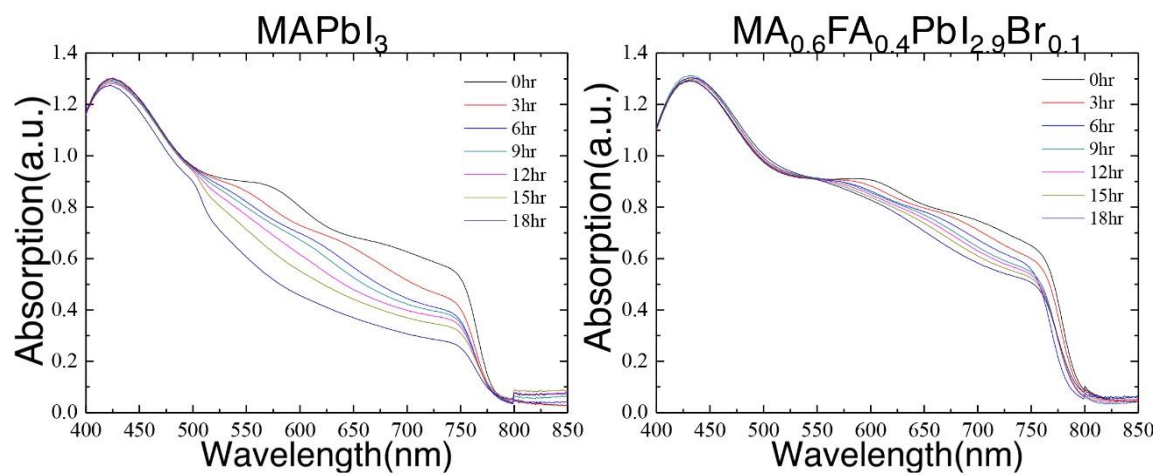

**Figure S8.** Comparison of the degradation rates of MAPbI$_3$ and MA$_{0.6}$FA$_{0.4}$PbI$_{2.9}$Br$_{0.1}$ perovskite films at 40% relative humidity with positive nitrogen ion deposition. Absorption spectra were measured at an interval of 3 hr.

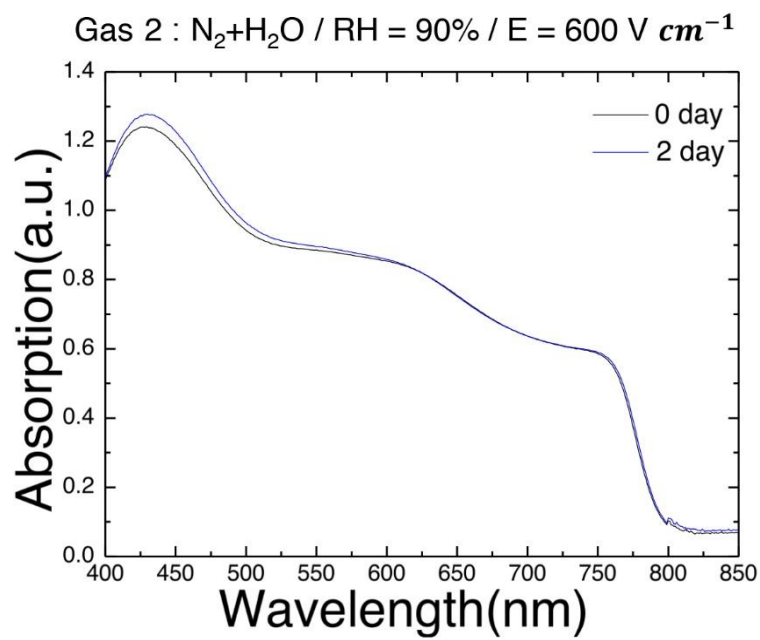

**Figure S9.** The effect of non-contact high electric field on the degradation of the perovskite film at high relative humidity (90%) under dark condition.

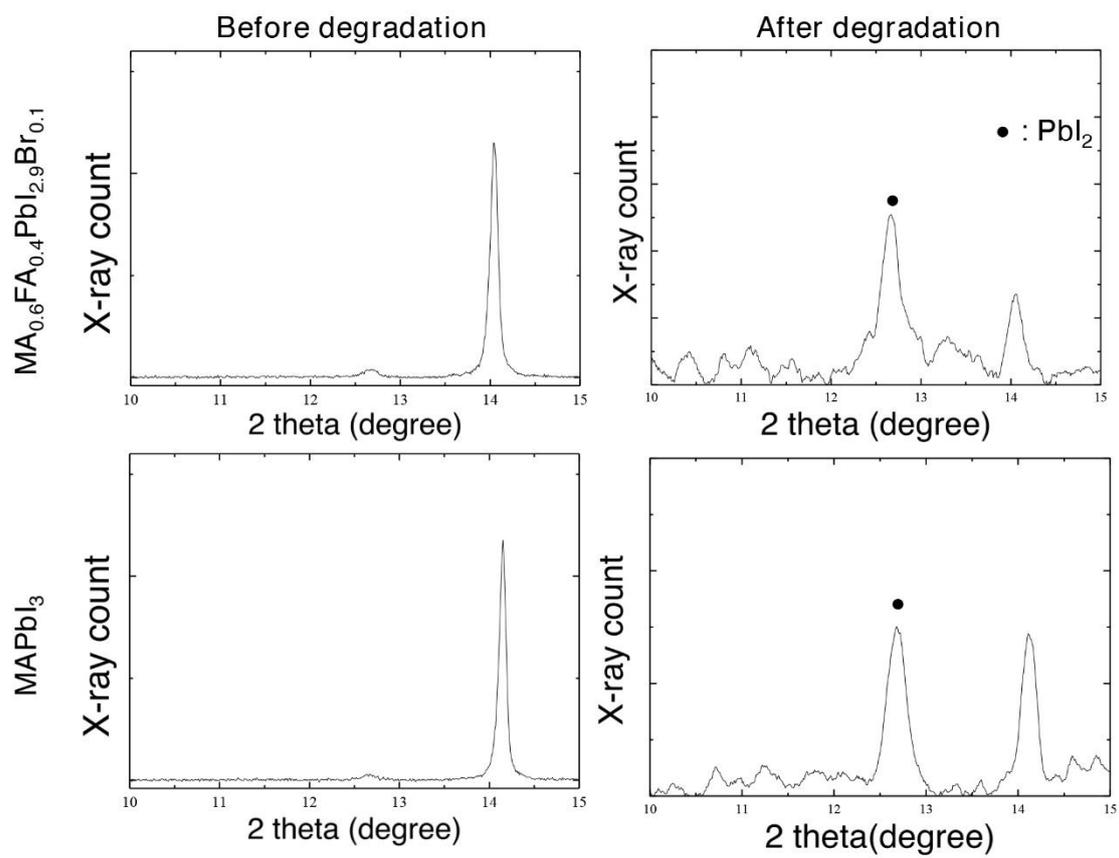

**Figure S10.** XRD patterns of the $MA_{0.6}FA_{0.4}PbI_{2.9}Br_{0.1}$ and $MAPbI_3$ film before and after degradation by trapped charges in the presence of moisture.

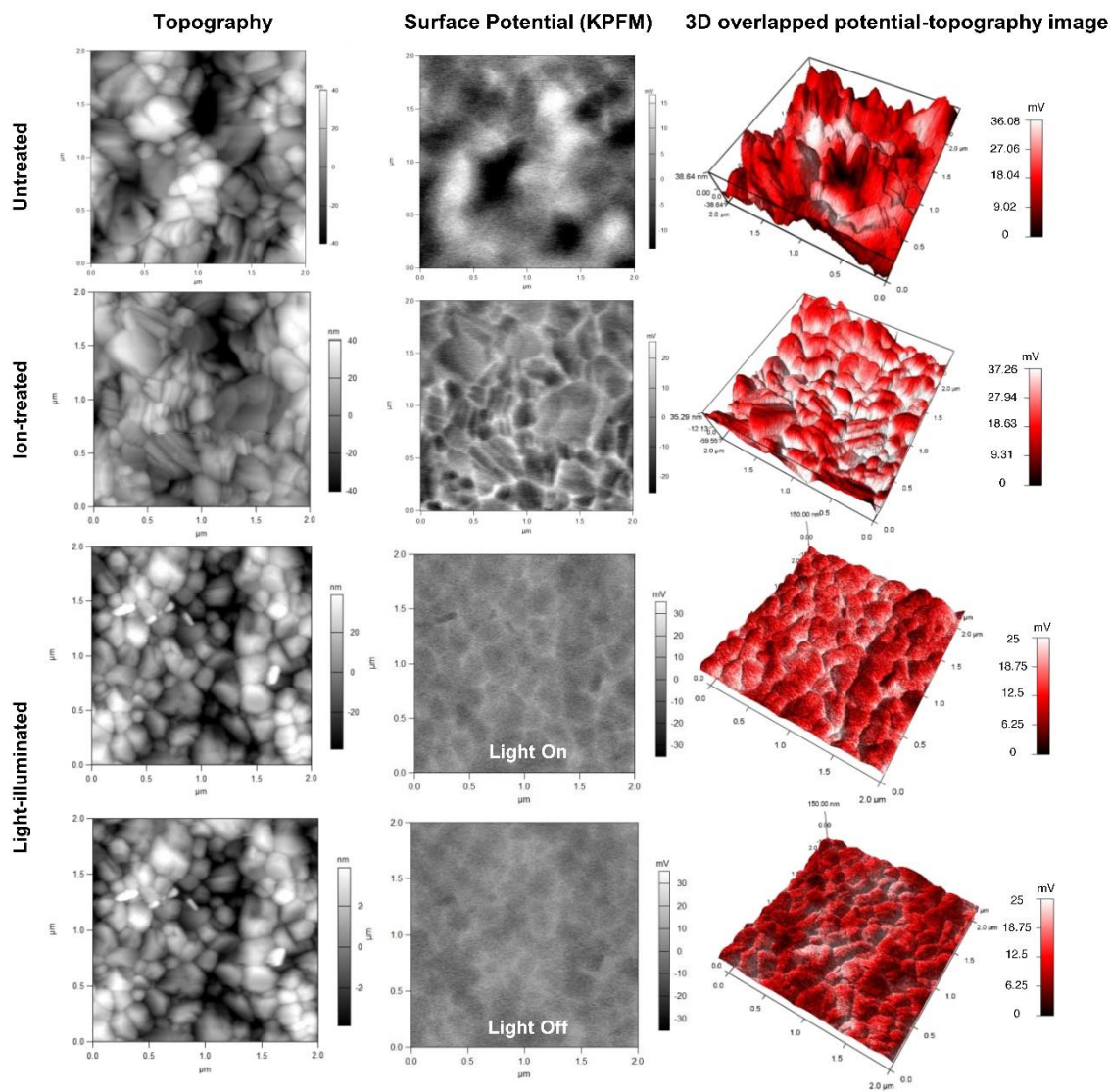

**Figure S11.** Topographies (first column) and surface charge density profiles (second column) of untreated (first row), Ion-treated (second row), and light-illuminated (third and fourth row) $MA_{0.6}FA_{0.4}PbI_{2.9}Br_{0.1}$ perovskite film. The images in the third and fourth row were obtained from light on and off during the measurement, respectively. The images in the third column show 3D plots of topographies colored based on the surface potential values. Both images of light illuminated cases show clear charge trapping along grain boundaries, but the charge trap is more contrasted when KPFM operation is under the light on

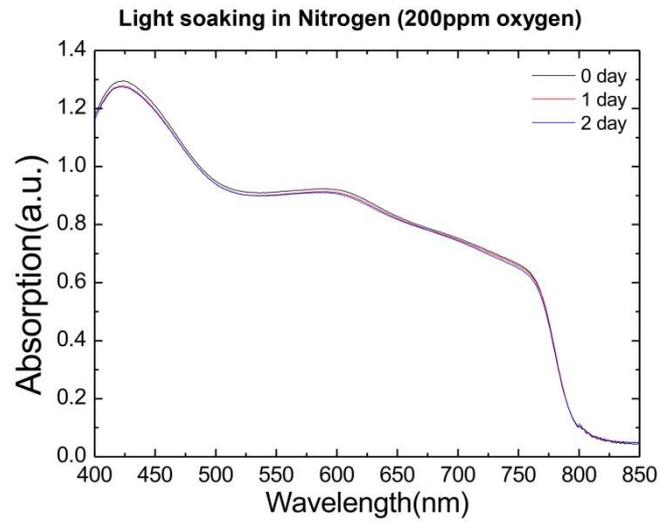

**Figure S12.** Absorption spectra of the perovskite film under one sun illumination in nitrogen. Light was illuminated in a moisture-free condition.

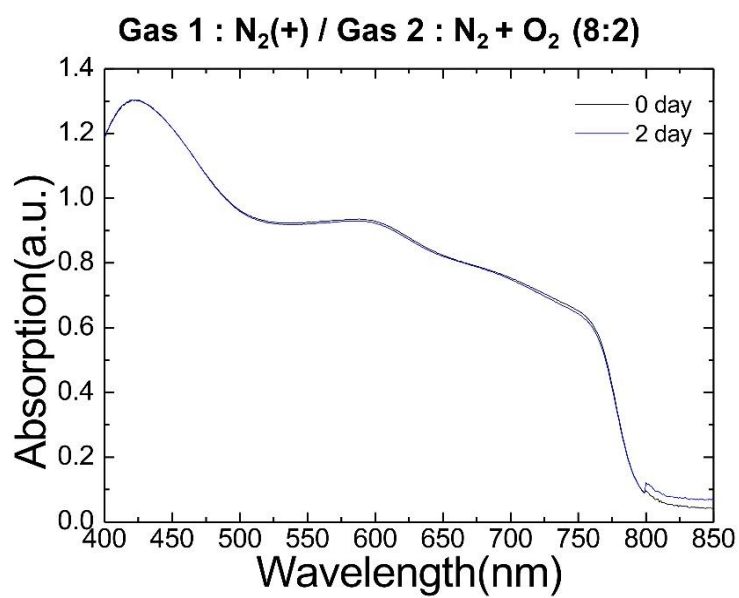

**Figure S13.** Absorption spectra of the perovskite film in dry air after deposition of positive nitrogen ions.